\documentclass[mathleft,fleqn,%
]{an}
\usepackage{graphicx}
\usepackage{subfig}
\usepackage[varg]{txfonts}
\usepackage{natbib}
\begin{document}

\Pagespan{1}{}
\Yearpublication{2016}%
\Yearsubmission{2015}%
\Month{0}%
\Volume{999}%
\Issue{0}%
\DOI{asna.201400000}%

\title{Carbon-enhanced metal-poor stars in different environments}  

\author{Stefania Salvadori\inst{1,2},\fnmsep\thanks{Corresponding author:
        {stefania.salvadori@obspm.fr}}
\'Asa Sk\'ulad\'ottir\inst{2}
\and  Matteo de Bennassuti\inst{3}
}
\titlerunning{Carbon-enhanced metal-poor stars in different environments}
\authorrunning{Salvadori, Skuladottir \& de Bennassuti}
\institute{
GEPI, Observatoire de Paris, PSL Research University, CNRS, Univ Paris Diderot, Sorbonne Paris Cité,  Place Jules Janssen, 92195 Meudon, France
\and 
Kapteyn Astronomical Institute, University of Groningen, Landleven 12 9747 AD,
Groningen, The Netherlands
\and 
INAF-Osservatorio Astronomico di Roma, Via di Frascati 33, I-00040 Monte 
Porzio Catone, Italy\\
}
\received{XXXX}
\accepted{XXXX}
\publonline{XXXX}
\abstract{The origin of carbon-enhanced metal-poor (CEMP) stars and their
possible connections with the chemical elements produced by the first stellar
generations is still highly debated. We briefly review observations of CEMP 
stars in different environments (Galactic stellar halo, ultra-faint and 
classical dwarf galaxies) and interpret their properties using cosmological 
chemical-evolution models for the formation of the Local Group. We discuss 
the implications of current observations for the properties of the first stars, 
clarify why the fraction of carbon-enhanced to carbon-normal stars varies in 
dwarf galaxies with different luminosity, and discuss the origin of the first 
CEMP(-no) star found in the Sculptor dwarf galaxy.}
\maketitle
\section{Background}
In the Local Group, spectroscopic studies of ancient individual stars, 
provide us with the unique opportunity to uncover the chemical enrichment 
of the interstellar medium when the Universe was less than 1 Gyr old. 
Thus, the fossil imprint of extinguished first stars can be found in 
these old stellar populations.\\ 
For more than a decade, the chemical signature of primordial pair 
instability supernovae (SN) with masses $M_*=(140-260) M_{\odot}$ 
(Heger \& Woosley 2002) have been looked for among the most metal-poor 
stars at [Fe/H]$<-3$, {\it in vain} (e.g. Cayrel et al. 2004). Still, 
state-of-the-art numerical simulations continue to predict that the 
first stars were likely very massive, $M_*=(10-1000)M_{\odot}$ (e.g. 
Hirano et al. 2014). Cosmological chemical evolution models for the 
Milky Way formation provide an explanation for such a tension between 
numerical findings and observations. ``Second-generation'' stars formed 
in environments polluted by pair instability SN {\it only}, are predicted 
to be extremely rare with respect to the overall Galactic halo population, 
and to have [Fe/H]$>-3$ if formed in small self-enriching galaxies 
(Salvadori et al. 2007; Karlsson et al. 2008).
The recent detection of a {\it rare} halo star at [Fe/H]$\approx-2.5$, 
likely showing the chemical signature of pair instability supernovae 
(Aoki et al. 2014), might be the first indication that this is the case. 
Thus, to catch these elusive relics we need to increase current stellar 
samples. Where we can find, instead, the chemical signature 
of less massive and less energetic first stars?\\
During the last few years an increasing number of carbon-enhanced metal-poor
(CEMP) stars, with [C/Fe]$ > 0.7$ and [Fe/H]$<-2$, have been found in the 
Galactic stellar halo and in nearby dwarf galaxies. The most metal-poor among 
them, at [Fe/H]$<-3$, do not typically show enhancement in slow (or rapid)
neutron capture elements produced by Asymptotic Giant Branch stars. Furthermore, 
they are not associated to binary systems, suggesting that their C-excess is 
likely representative of their environment of formation (Norris et al. 2010). 
These ``CEMP-no'' stars become more frequent as we move towards lower [Fe/H], 
and their C-excess gradually increases (Fig.~1). CEMP-no stars at [Fe/H]$<-5$ 
show very peculiar chemical abundance patterns, which are consistent with birth 
environments polluted by $M_*=(10-40)M_{\odot}$ {\it primordial} faint SN that 
developed mixing and fallback (e.g. Iwamoto et al. 2005).
A relatively good agreement with data is also obtained by models of {\it primordial} 
(or low-metallicity) ''spinstars'', $M_*=(40-120)M_{\odot}$, that experience mixing 
and mass loss because of their high rotational velocity (e.g. Meynet et al. 2006).
In conclusion, available observations support the idea of a link between CEMP-no 
stars and moderately massive first stars.
\section{Observations and open issues}
In Fig.~1 we show a collection of Carbon measurements in metal-poor stars dwelling 
in the oldest component of the Local Group: the Galactic stellar halo, ultra-faint 
dwarf galaxies, and the dwarf spheroidal galaxy Sculptor. 
According to hierarchical structure formation models, dwarf galaxies are expected 
to be the building blocks of stellar haloes (e.g. Helmi et al. 2008). Thus we are 
comparing systems that likely experienced similar early star-formation histories.
In Fig.~1 (left) we can first note that in the stellar halo C-enhanced and C-normal 
stars already co-exist at [Fe/H]$\approx -4.75$. A key question is then: what 
physical mechanism determines the formation of these different classes of stars? 
Are all CEMP-no stars second-generation objects?\\  
We can also note that many CEMP-no stars have been found in ultra-faint dwarf 
galaxies, the faintest and most metal-poor satellites of the Milky Way, which 
have total luminosities $L<10^5L_{\odot}$. In these systems, the [C/Fe] vs [Fe/H] 
measurements follow the same trend observed in the Galactic halo, but the fraction 
of CEMP-no stars at [Fe/H]$<-3$ with respect to the total is even {\it higher} 
(Salvadori et al. 2015). Deep color-magnitude diagrams of ultra-faint dwarfs show 
that these galaxies are dominated by $>12$~Gyr old stars (Brown et al. 2014), 
confirming that they might be the living fossils of the first galaxies, which 
formed prior the end of reionization and hosted the first stars (Bovill \& 
Ricotti 2009; Salvadori \& Ferrara 2009).\\
In the more luminous ``classical'' dwarf spheroidal galaxy Sculptor ($L\approx 
10^{6.3}L_{\odot}$) instead, CEMP-no stars are much more rare than in the Galactic 
halo and in ultra-faint dwarfs (Fig.~1, right). In spite of very accurate and 
intense searches, no CEMP stars have been (yet) found at the lowest [Fe/H] (see 
Sk\'ulad\'ottir et al. 2015). This result is quite surprising since Sculptor is 
also dominated by an old stellar population, and thus first stars are expected 
to be formed in this galaxy. If CEMP-no stars trace the early chemical enrichment 
by the first stars, why they are not observed in Sculptor? The puzzle became 
even more intricate after the discovery of first CEMP-no star in Sculptor 
(Sk\'ulad\'ottir et al. 2015). In fact, this star has an {\it unusually high} 
[Fe/H]$\approx -2$ (Fig.~1, right) although its chemical abundance pattern is 
consistent with an environment of formation also enriched by primordial faint 
SN (Sk\'ulad\'ottir et al. 2015). This observation poses several questions: 
does the fraction of CEMP-no stars depend on galaxy luminosity? Are CEMP-no 
stars at lower [Fe/H] absent or hidden in Sculptor?
\begin{figure*}[!tbp]
  \centering
  \subfloat{\includegraphics[width=0.45\textwidth]{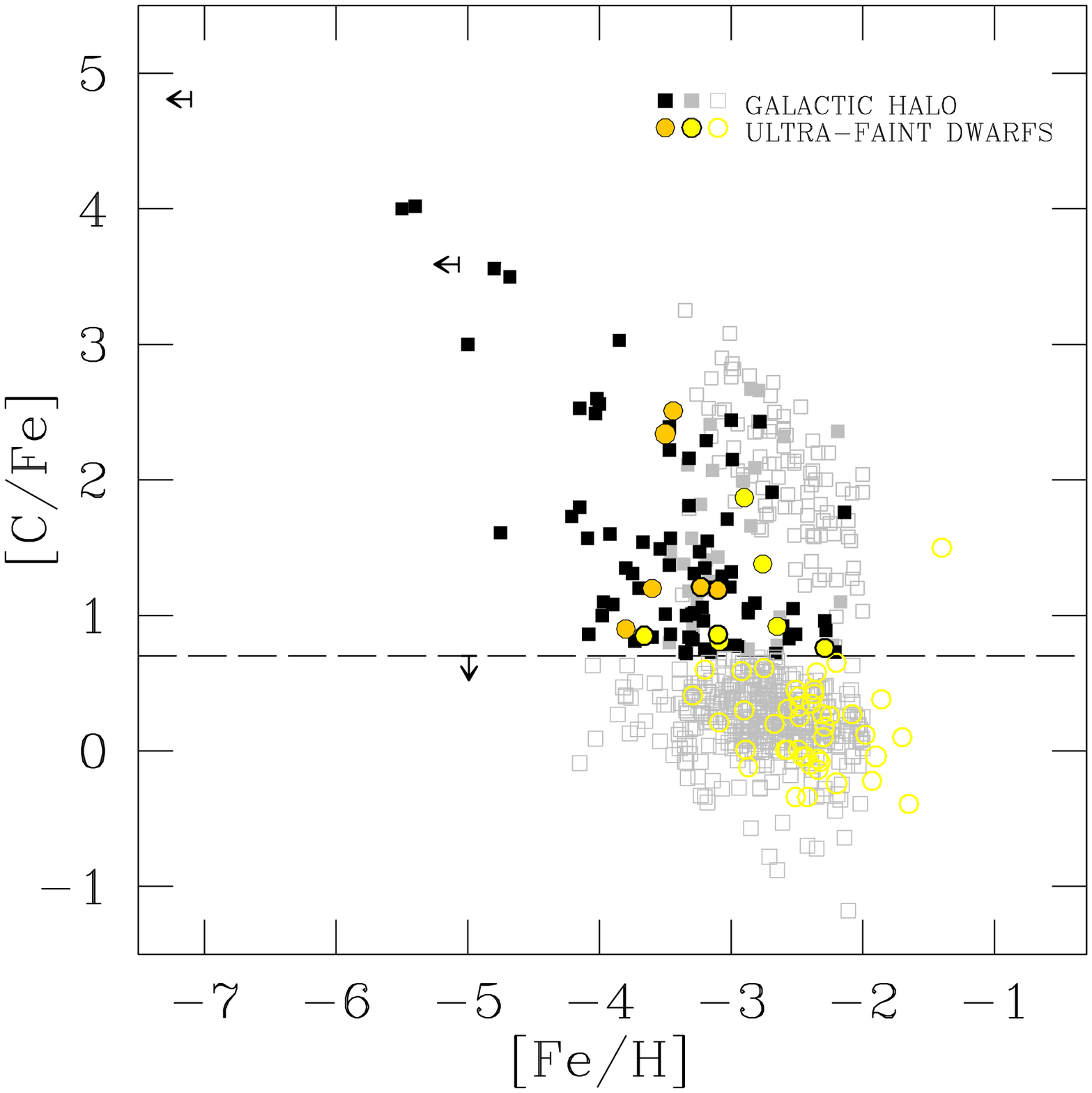}\label{FIG1a}}
  \hfill
  \subfloat{\includegraphics[width=0.45\textwidth]{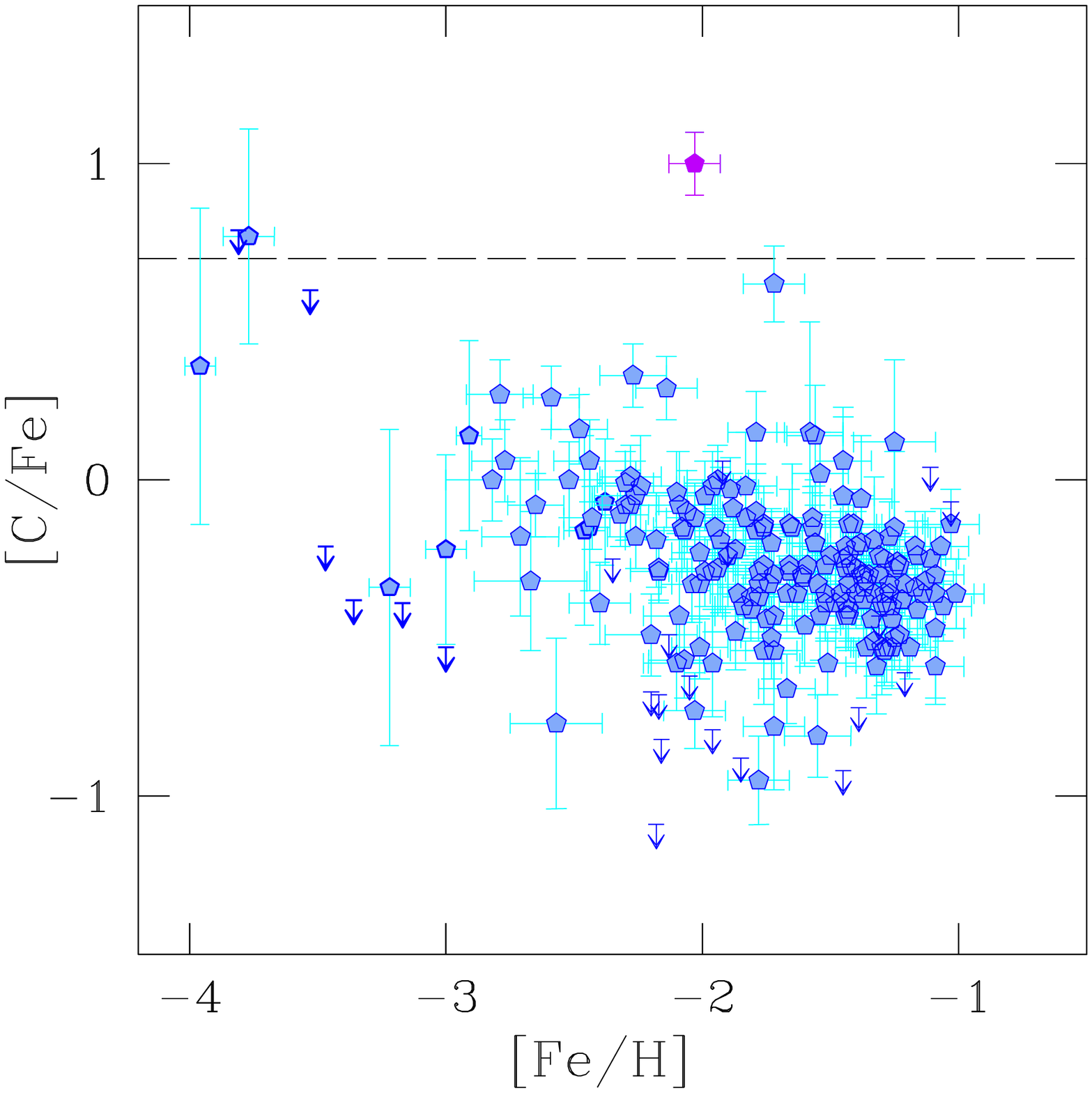}\label{fig1b}}
  \caption{Compilation of stars with measured [C/Fe] and [Fe/H] - see 
     Fig.~1 of \cite{salvadori15} for references.
     {\it Left}: stars in the Galactic halo ({\it squares}) and ultra-faint 
     dwarfs ({\it circles}). CEMP-no stars are shown with dark filled symbols 
     ({\it black, orange}), light filled symbols show CEMP stars with no 
     available measurements of slow (rapid) neutron capture elements, and 
     open symbols all the remaining stars. {\it Right}: C and Fe measurements
     for stars in the Sculptor dwarf galaxy.}
\end{figure*}
\section{How do we interpret current observations?}
To interpret these observations in terms of primordial cosmic star-formation 
and early galaxy evolution, we can use cosmological models for the build-up 
of the Local Group. Such a statistical tools, which catch the essential physics 
of early galaxy formation, follow the star-formation and chemical evolution 
of the Milky Way (Salvadori et al. 2007; de Bennassuti et al. 2014) and nearby 
dwarf galaxies (Salvadori \& Ferrara 2009; Salvadori et al. 2015) along their 
possible merger histories by resolving star-formation in $10^{6.5}M_{\odot}$ 
mini-haloes (Salvadori \& Ferrara 2009).
The minimum halo mass to form stars is assumed to increase with cosmic time 
to account for the gradual effect of reionization (Salvadori et al. 2014). 
First stars with a variable mass distribution are assumed to form when the 
amount of dust and metals in the star-forming gas is lower than the critical 
value to allow gas fragmentation, $Z_{cr}<10^{-4}Z_{\odot}$ (de Bennassuti 
et al. 2014).
Otherwise, ``normal'' stars form according to a Larson initial mass function 
(IMF). Stars with different masses contribute to the chemical enrichment in 
their proper time scales, and SN explosions are assumed to eject metals and 
gas into the surrounding Milky Way environment, where the heavy elements gets 
instantaneously mixed (see Salvadori et al. 2014 for the inhomogeneous metal 
enrichment treatment). Both the efficiency of star-formation and the SN winds 
are fixed to reproduce the global properties of the Milky Way and they are 
assumed to be the {\it same} for all star-forming haloes. The only exceptions 
are mini-haloes, in which the star-formation efficiency is supposed to be 
reduced to account for ineffective cooling by molecular hydrogen (Salvadori 
\& Ferrara 2009; Salvadori et al. 2015). 
For the details we remind the reader to the original papers.
\begin{figure}[!tbp]
  \includegraphics[trim=1.5cm 2.cm 9.cm 1.0cm, clip=true,width=7.1cm]{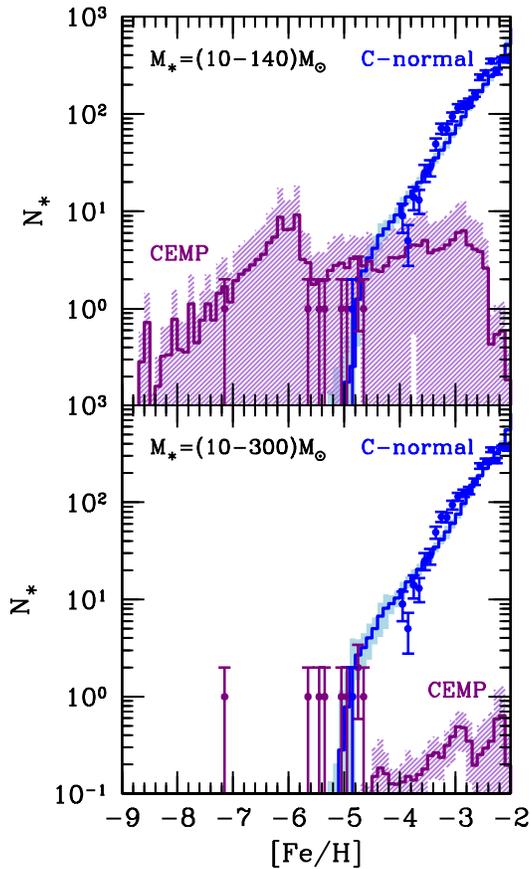}
  \caption{The observed ({\it points}) and simulated ({\it histograms}) 
     MDFs for Galactic halo stars obtained by assuming different mass ranges 
     for the first stars (see labels). We show C-enhanced ({\it violet}) and 
     C-normal stars ({\it blue}). Shaded regions are $\pm 1\sigma$ dispersion 
     among 50 Milky Way merger histories.}    
\end{figure}

Let's first focus on Galactic halo stars. In Fig.~2 we compare the observed 
Metallicity Distribution Function (MDF) with model results obtained by assuming 
that first stars form according to a Larson IMF with different mass ranges: 
$M_*=(10-140)M_{\odot}$ (top) and $M_*=(10-1000)M_{\odot}$ (bottom). We can 
immediately see that the low-Fe tail of the MDF, which is populated by CEMP-no 
stars, is extremely sensitive to the assumed mass range of the first stars 
(de Bennassuti et al. 2014). Our models show that a good match
to the observations requires $M_*=(10-140)M_{\odot}$ (Fig.~2 top). This means 
that the early metal-enrichment should be dominated by primordial {\it faint} 
SN, which have masses $M_*\approx (10-40)M_{\odot}$ and produce large amounts 
of C and very small of Fe (e.g. Iwamoto et al. 2005). When the contribution from 
energetic pair instability SN is also accounted for (Fig~2, bottom), the chemical 
signature of faint SN is completely washed out and CEMP-no stars at 
[Fe/H]$<-4.5$ 
are not predicted to exist. According to our findings CEMP-no stars at [Fe/H]$<-5$ 
are truly second-generation objects, which have been enriched by primordial faint SN {\it only}. As we move towards higher [Fe/H], CEMP-no 
stars form in environments polluted by both primordial faint SN and normal type 
II SN, which start to dominate the chemical enrichment at very high 
redshifts (Salvadori et al. 2014). Normal SN typeII are the main pollutants of the inter-stellar medium of formation 
of C-normal stars, which are predicted to exist already at [Fe/H]$\approx -4.75$ 
(Fig. 2; de Bennassuti et al. 2014; Salvadori et al. 2015). 
In conclusion, the existence of CEMP-no stars at [Fe/H]$<-4$ provide key 
information on the mass range of the first stars, suggesting that faint 
primordial SN, with $M_*=(10-40) M_{\odot}$, must have dominated the early 
phases of chemical evolution.\\
Thus, to investigate the incidence of CEMP-no stars in dwarf galaxies with 
different luminosities, we can simply assume that the first stars have all 
masses $M_*=25M_{\odot}$ and evolve as faint SN (Salvadori et al. 2015). With 
this working-hypothesis we find that, at [Fe/H]$< -3$, the average fraction 
of CEMP-no stars with respect to the total follows almost the same trend in 
all dwarf galaxies (see Fig.~3 of Salvadori et al. 2015). This ``universal'' 
shape is a consequence of the underlying hierarchical $\Lambda$CDM model for 
structure formation, according to which all galaxies built-up through merging 
of progenitor mini-haloes (e.g. Salvadori et al. 2010), where [Fe/H]$<-3$ 
 stars predominantly form. 
In spite of that, we find that the probability to 
observe a CEMP-no star in a given [Fe/H] range strongly depends on the galaxy 
luminosity and it is one order of magnitude higher in ultra-faint 
dwarfs than in the Sculptor dwarf galaxies (Salvadori et al. 2015). 
This is due to the dramatic change, with increasing luminosity, of the MDF of 
dwarf galaxies as shown in Fig.~3. We can see that, on average, the MDFs of 
ultra-faint dwarfs cover a broader [Fe/H] range than Sculptor-like dwarfs. 
Furthermore, they are flatter, and thus contain more stars at [Fe/H]$<-3$, 
where CEMP-no mostly reside. Such a shape is a consequence of the low 
star-formation rate of ultra-faint dwarfs, which are predicted to be associated 
to low-mass mini-haloes (Salvadori \& Ferrara 2009; Salvadori et al. 2015). 
More luminous dwarf galaxies, instead, result from the merging of these small 
systems and more massive progenitors, which assembled at later epochs from 
metal enriched regions of the Milky Way environment, and have higher star-formation 
efficiencies. Their MDFs are hence peaked and shifted towards higher [Fe/H],
where CEMP-no stars can be more likely found.\\ 
Consequently, the fraction of stars at [Fe/H]$<-3$ with respect to the 
total dramatically decreases as the galaxy luminosity increases (Fig.~3). 
In Sculptor-like dwarf galaxies we find that these extremely metal-poor
stars only represent $<3\%$ of the total. Thus, we predict that CEMP-no 
stars at [Fe/H]$<-3$ are not lacking in Sculptor but they are {\it hidden} and 
thus more difficult to catch. We can then ask: how much should we enlarge 
the current stellar sample to uncover the low-Fe tail of the Sculptor MDF?
\begin{figure}[h]
\includegraphics[width=\linewidth]{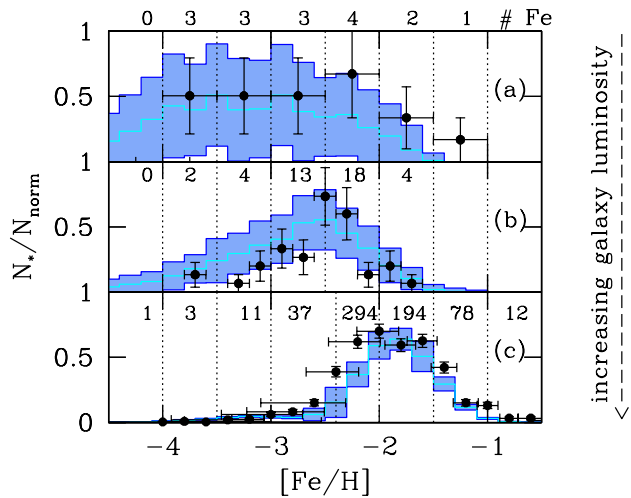}
\caption{Observed ({\it points}) and predicted ({\it histograms}) MDFs of: 
     (a) ultra-faint dwarf galaxies with $L<10^4L_{\odot}$;
     (b) Bootes-, $L\approx 10^4 L_{\odot}$, and 
     (c) Sculptor-like dwarf galaxies, $L\approx 10^6 L_{\odot}$. 
     Labels indicate the number of measurements in each [Fe/H] bin.
     We show the $\pm1\sigma$ dispersion among 100 Monte Carlo 
     sampling of the average MDF to the number of stars observed (see
     Salvadori et al. 2015).}     
\label{fig3}
\end{figure}
\section{Looking ahead: model predictions}
\begin{figure}[h]
\includegraphics[width=0.93\linewidth]{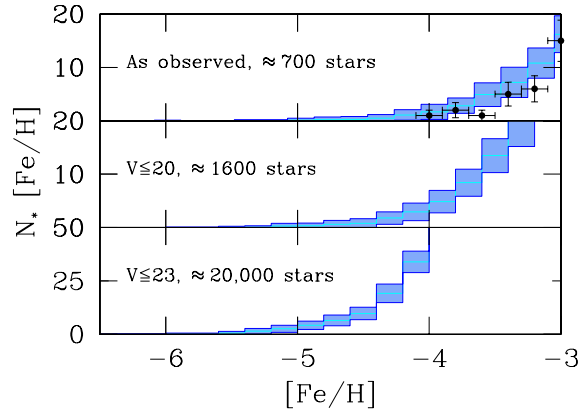}
\caption{Number of stars at [Fe/H]$<-3$ that are predicted
     to be observed in Sculptor by increasing the [Fe/H] 
     measurements. From top to bottom we show results for:
     i) the current statistics ii) stars with $V\leq20$, 
     and iii) $V\leq 23$. Shaded area show the $\pm 1\sigma$ 
     errors (see Fig.~3).} 
\label{fig4}
\end{figure}
Fig.~4 shows how many stars at [Fe/H]$<-3$ will appear in Sculptor by 
targeting fainter stars and thus increasing the overall sample of [Fe/H] 
measurements. By using current facilities, we can follow-up stars down 
to $V\leq 20$. This might allow us to observe $12\pm 8$ stars with [Fe/H]$<-4$, 
the $\approx 40\%$ of which should be CEMP-no stars. With new generation 
instruments and telescopes, such as MOSAIC on the ESO-Extremely Large Telescope, we will be able to dramatically increase the statistics by 
observing all the stars in Sculptor down to the main sequence turn-off 
(Evans et al. 2015). 
In this case, we might be able to catch $\approx 80$ stars at [Fe/H]$<-4$ 
and $\approx 16$ at [Fe/H]$<-4.7$, where the incidence of CEMP-no stars 
should be $100\%$ (Fig.~4c, see Salvadori et al. 2015 for details). 
These experiments, therefore, will allow us to test the predominant role 
of faint primordial SN on early metal enrichment, along with the underlying 
hierarchical models for structure formation.\\
In conclusion, Near-Field cosmology is a powerful strategy to (indirectly) 
study the properties of extinguished first stars and the physics of early 
galaxy formation. Larger stellar samples are required to tightly constrain 
the mass spectrum of the very first stars. 
In the nearby future, we will have at our disposal results from wide 
and deep spectroscopic surveys, such as those discussed in this volume. 
By combining theoretical and observational efforts we can exploit these 
data to unveil the first star properties. Thus, we are entering in the 
Golden-Era of Near-Field cosmology.
\acknowledgements
  We are in debt with E. Tolstoy and R. Schneider. SS thanks the 
  conference organizers for a very productive meeting. 
  She is grateful to NWO for her VENI grant: 639.041.233.

\end{document}